# Wavelength flattened directional couplers for mirror-symmetric interferometers

**Maurizio Tormen and Matteo Cherchi**

Pirelli Labs – Optical Innovation, viale Sarca 222, 20126 Milano, Italy

Abstract: In the context of guided optics, we derive, analytically and geometrically, a rigorous general criterion to design wavelength insensitive interferometers with mirror symmetry, which are needed for wavelength multiplexing/demultiplexing. The criterion is applied to a practical case, resulting in an interferometer that works on a band wider than 70 nm.

**Introduction**

Optical waveguides and directional couplers are the basic elements of planar lightwave circuits. Unlike their free space counterparts, they confine light on cross sections comparable with the square of the wavelength. As a consequence their response will be, in general, strongly wavelength dependent. This is why many different wavelength flattened directional couplers[1-9] have been proposed. In particular they can be used as 50-50 splitters (see Fig. 1) in Mach-Zehnder interferometers[10] (MZI), where, for most applications, their response must be flat over the whole band of interest.

In the context of wavelength multiplexing/demultiplexing the MZI is commonly used in combination with Bragg gratings[11,12] to get 2x2 port add and drop filters (Fig. 2). Two identical gratings are placed symmetrically on both arms, between the couplers of the interferometer, acting as wavelength selective elements. They can be designed as passband filters, stopband filters, band splitters, etc.. The structure acts as a standard MZI for all transmitted wavelengths, whereas it is equivalent to a Michelson interferometer for all reflected wavelengths. Clearly, for telecom applications, both transmitted and reflected power should be addressed in the cross port, i.e. the port other than the input port. In other words, referring to Fig. 2b, transmitted input power must be addressed in the thru port together with reflected added power as well as reflected input power must be addressed in the drop port. For DWDM applications at least 30 nm wide bands are required and the interferometer must have a flat response over the whole band of interest, both in transmission and in reflection. This is even more important when dealing with tunable gratings[13]. In this paper we will show that it is not trivial to design a wavelength insensitive Michelson interferometer even when starting from a wavelength flattened directional coupler.

## 50-50 Couplers and MZI Symmetry

We will assume a power splitter as a lossless device made of two waveguides, one beside the other along the propagation axis $z$. In general their cross-sections, their distance and their refractive index profiles may vary along $z$. We will also assume that, for any fixed $z_0$, the two eigenmodes of the single waveguide cross-section, $E_j(z_0) \equiv E_j(x, y, z_0)$, ($j=1,2$) can be considered as a basis for the field at that point, so that it is always possible to write:

$$E(z) \equiv a_1(z) \, E_1(z) + a_2(z) \, E_2(z)$$

where $a_j(z)$ are complex numbers so that $|a_1(z)|^2 + |a_2(z)|^2 = 1$ and $E_j(z)$ are normalized so that $|a_j|^2$ represent the power fractions in each of the waveguides. We will regard $E(z)$ as a scalar

quantity, since we will consider only singly copolarized modes of the structure. With these assumptions we can always represent the generic state of the system through the complex vector

$$\mathbf{u}(z) \equiv \begin{pmatrix} a_1(z) \\ a_2(z) \end{pmatrix}. \tag{1}$$

Its evolution can be described by an unitary 2x2 matrix or, up to an overall phase, by an element $\mathbf{U}$ of $SU(2)$[14] (i.e. a matrix so that $\mathbf{U}^{-1} = \mathbf{U}^\dagger$ and $\det \mathbf{U} = 1$). From its definition, it can be always cast in the form

$$\mathbf{U} = \begin{pmatrix} \exp(-i\theta/2)\cos\phi & \exp[i(\eta+\theta/2)]\sin\phi \\ -\exp[-i(\eta+\theta/2)]\sin\phi & \exp(i\theta/2)\cos\phi \end{pmatrix}. \tag{2}$$

If optical power is launched in any of the four ports, the splitting ratio (defined as the power fraction transferred to the other waveguide) does not depend on which port is chosen and it is clearly given by $\sin^2\phi$.

We will call "50-50 splitter" any combination of phase shifters and synchronous, asynchronous or tapered couplers giving 50-50 power splitting. The whole 50-50 transformation can be always represented, up to an overall phase, as

$$\mathbf{S} = \frac{1}{\sqrt{2}} \begin{pmatrix} \exp(-i\theta/2) & \exp[i(\eta+\theta/2)] \\ -\exp[-i(\eta+\theta/2)] & \exp(i\theta/2) \end{pmatrix}. \tag{3}$$

We will also assume, without loss of generality, that a MZI is simply a cascade of two 50-50 splitters, since any phase shift can be seen as part of the splitters.

A MZI composed of two identical splitters (Fig. 1.a) in general doesn't guarantee full power transfer in one of the two output ports. In particular full power transfer in the cross port can be achieved if and only if the transformation is of the form

$$\overline{\mathbf{S}} = \frac{1}{\sqrt{2}} \begin{pmatrix} 1 & e^{i\eta} \\ -e^{-i\eta} & 1 \end{pmatrix}. \quad (4)$$

For example a 50-50 synchronous coupler belongs to this class and corresponds to $\eta = \pi/2 + k\pi$ with $k$ integer.

More remarkably when any 50-50 splitter is cascaded with its point-symmetry transformed (Fig. 1.b), power is always transferred in the cross port. This is clear when considering that the point-symmetric splitter is represented by the matrix[10] (obtained by exchanging the diagonal elements of $\mathbf{S}$)

$$\mathbf{S}_p \equiv \frac{1}{\sqrt{2}} \begin{pmatrix} \exp(i\theta/2) & \exp[i(\eta+\theta/2)] \\ -\exp[-i(\eta+\theta/2)] & \exp(-i\theta/2) \end{pmatrix} \quad (5)$$

that gives $(\mathbf{S}_p\mathbf{S})_{11} = (\mathbf{S}_p\mathbf{S})_{22}^* = 0$.

This is a very important result because it means that, given a wavelength flattened 50-50 splitter, it is always possible to get a wavelength insensitive MZI by simply choosing a point-symmetric configuration. Notice that also the form $\overline{\mathbf{S}}\overline{\mathbf{S}}$ of Eq. 4 belongs to the class of point-symmetric configurations, since it satisfies $\overline{\mathbf{S}}_p = \overline{\mathbf{S}}$. This means that the single 50-50 splitter is itself point-symmetric. Actually it is easily shown that $\mathbf{S} = \mathbf{S}_p$ if and only if $\mathbf{S}$ is of the form $\overline{\mathbf{S}}$, that is cascading two identical 50-50 splitters we can get full power transfer in the cross port if and only if the single splitter is itself point-symmetric.

Anyway, if we consider a grating on a MZI, the point symmetric configuration can guarantee a proper response for transmitted wavelengths, but not for reflected wavelengths, since they pass again in the first splitter. This is equivalent to be launched in a mirror-symmetric replica of the same splitter. Therefore, if we cascade to a first splitter its mirror symmetric replica (Fig. 2.b), both transmitted and reflected wavelengths will experience the same type of interferometer. The building block of this kind of structure must be a 50-50 splitter that works when cascaded with its mirror-symmetry transformed, that matricially is represented by the transposed[10] of $\mathbf{S}$

$$\mathbf{S}_m \equiv \mathbf{S}^t = \frac{1}{\sqrt{2}} \begin{pmatrix} \exp(-i\theta/2) & -\exp[-i(\eta+\theta/2)] \\ \exp[i(\eta+\theta/2)] & \exp(i\theta/2) \end{pmatrix}. \tag{6}$$

Imposing full power transfer on the cross port, we find that $\mathbf{S}$ must be of the form

$$\tilde{\mathbf{S}} = \frac{1}{\sqrt{2}} \begin{pmatrix} \exp(-i\theta/2) & \pm i\exp(i\theta/2) \\ \pm i\exp(-i\theta/2) & \exp(i\theta/2) \end{pmatrix}. \tag{7}$$

In order to better understand the physical meaning of this form we will now introduce a convenient geometric representation of the $\mathbf{S}$ transformations.

## Geometric Representation

It is well known[14-18] that SU(2) transformations may be mapped into SO(3) transformations through a homomorphism. This means that all the transformation we have analyzed before can be represented as rotations on a spherical surface, analogous of the Poincaré sphere for polarization states. In Fig. 3 are displayed all the intersections of the $S_1$, $S_2$, $S_3$ axes with the sphere. They represent respectively the single waveguides modes $E_1$, $E_2$ and their linear combinations $E_{S,A} \equiv \frac{1}{\sqrt{2}}(E_1 \pm E_2)$ and $E_{R,L} \equiv \frac{1}{\sqrt{2}}(E_1 \pm i E_2)$. Also the generic normalized state $P \equiv a_1 E_1 + a_2 E_2 \equiv \cos\alpha\, E_1 + \exp(i\theta)\sin\alpha\, E_2$ is plotted. From this expression it is clear that the circle in the $S_2 S_3$ plane is the locus of 50-50 power splitting. When considering transformations which are compositions of reciprocal elements only (which is the case when dealing with directional couplers and phase shifters), all trajectories on the sphere can be compositions of rotations about axes belonging to the equatorial plane only[18], i.e. the $S_1 S_2$ plane. In particular the action of a phase shifter is represented by a rotation about the $S_1$ axis, and the action of a synchronous coupler is represented by a rotation about the $S_2$ axis. Rotation axes between $S_1$ and $S_2$ correspond to asynchronous couplers, that can be seen as combinations of couplers and phase shifters. This means that the projection of any physical trajectory on the $S_1 S_2$ plane will be always a composition of straight lines. Clearly the projection on the same plane of the 50-50 splitting locus coincide with the $S_2$ axis.

We are now able to show the following

*Criterion*: if a 50-50 splitter is cascaded to its mirror-symmetric, the necessary and sufficient condition to get full power transfer in the cross port is that the first splitter sends $E_1$ into the state $E_R$ or into the state $E_L$.

Consider the form $\tilde{\mathbf{S}}$ in Eq. 7. It sends the input states $E_1$ or $E_2$ into the states $E_R$ or $E_L$, that analytically prove the necessity. To show the sufficiency, notice that the projection on the $S_1 S_2$ plane of $E_R$ or $E_L$ is the center $O$ of the equator. Figure 4.a shows the projection $\Gamma$ on the equatorial plane of a generic trajectory going from $E_1$ to, say, $E_R$. When starting from $O$, the projection $\Gamma_m$ of the trajectory associated to the mirror-symmetric system will be the point-symmetric of $\Gamma$ with respect to $O$ (mirror symmetry leaves the rotation axes unchanged and inverts the order of the transformations only). Since the point symmetric of $E_1$ with respect to $O$ is $E_2$, this geometrically prove the sufficiency of the condition. This criterion fully determine the class of 50-50 splitters that work in the Michelson configuration.

In Fig. 4.b we have also shown a geometrical interpretation of the previous result about point-symmetric structures. The projection $\Gamma$ of the trajectory associated with a generic 50-50 splitter will reach a generic point on the $S_2$ axis. Starting from this point, the projection $\Gamma_p$ of the trajectory associated to the point-symmetric system will be the mirror-symmetric of $\Gamma$ with respect to the $S_2$ axis (point-symmetry not only inverts the order of the transformations, but also changes sign to the phase shifts, i.e. rotates about the mirror-symmetric axes) and will reach the mirror-symmetric of $E_1$, that is $E_2$.

Notice that the point-symmetric configuration works for any 50-50 splitter, whereas the criterion for mirror-symmetric structures imposes also a fixed phase relation. This means that, to get wavelength insensitive interferometers, in the first case we just need wavelength flattened power splitting, while in the second case we also need wavelength independent phase response, that is to flatten the relative phase.

## 50-50 Mach-Zehnder Flattened Splitters

A very effective wavelength flattened 50-50 splitter is the Mach-Zehnder Coupler[5-8] (MZC). It is composed by a ±120° phase shifter between a "half" coupler, that is a 50-50 splitter composed by a single synchronous directional coupler, and a "full" coupler, i.e. a coupler that is the double of the previous one. In a recent paper we have also geometrically interpreted its working principle[18]. Figure 5.a shows the trajectory on the sphere corresponding to a Full-Half (FH) configuration with a -120° phase shift, when the wavelength dependence makes the couplers non ideal. The same is plotted in Fig. 5.b for the Half-Full (HF) configuration. In both configurations the trajectory ends on the 50-50 splitting circle. In the FH configuration close to the ideal case ending point $E_L$, and in the HF configuration close to the ideal case ending point, corresponding to 30° relative phase. In the non-ideal case (that is at a wavelength different by the nominal wavelength) to the half coupler and to the full coupler correspond respectively a $\Delta\theta$ and a $2\Delta\theta$ coupling angle error.

In the FH configuration, before the phase shift the trajectory has an inclination of $2\Delta\theta$ over the $S_1S_2$ plane, away from the ideal ending point, whereas after the phase shift (remember that cos(120°) = -1/2) the inclination is $\Delta\theta$ below the same plane, i.e. closer to the ideal ending 50-50 splitting circle. So the -120° phase shift compensates at once the error of both couplers, as clearly shown in Fig 5.a.

In the HF configuration the half coupler trajectory ends at a distance $d = \Delta\theta \cdot R$ from the 50-50 splitting circle (being $R$ the sphere radius). After a -120° phase shift, the full coupler trajectory is along a R/2 circle so that a $d = 2\Delta\theta \cdot R/2$ error is fully corrected, as clearly shown in Fig. 5.b.

We also notice that in the FH configuration the change of the ending point strongly depends on the error in the coupling angle, whereas in the HF configuration this change is very small. In fact in the first case the departure from $E_L$ is proportional to the change of the radius of the phase shifter rotation, which, in turn, is directly proportional to the coupler error. On the contrary, in the second case the radius of the phase shifter rotation is only slightly affected by the coupling change, resulting in a negligible departure from the ideal 30° phase shift point. This is quantitatively shown in Fig. 6, a transfer matrix simulation of a MZC which takes in account the wavelength dependence of couplers and phase shifters as calculated with a commercial vectorial mode-solver[19]. We have considered glass waveguides with 4.5% index contrast, square 2μm wide cross section and couplers with 2μm minimum inner wall separation. It is clearly shown that, when departing from the nominal wavelength $\lambda_0 = 1310\text{nm}$, both configurations have a flattened splitting ratio, but only the second one features a flattened phase response.

## Mach-Zehnder/Michelson Interferometers

We can now apply the criterion for mirror-symmetric interferometers to the wavelength flattened power splitter in order to get wavelength insensitive Michelson interferometers.

### *HF configuration*

The phase stability of the HF MZC makes it the natural candidate to fulfil the criterion we have previously derived for mirror symmetric interferometers. In fact we have just to change the 30° output relative phase with an additional 60° phase shift in order to reach $E_R$, as required by the criterion (Fig. 7.a). Power and phase flatness of the 50-50 splitter together with wavelength stability of first order phase shifters ensure wavelength insensitivity of the resulting interferometer, both in reflection and in transmission.

### *FH configuration*

Even though the FH MZC does not feature a stable phase output, this limitation can be overcome with a convenient wavelength dependent phase shifter. Let us define $\phi(\lambda)$ the output relative phase of the MZC that, at the nominal wavelength $\lambda_0 = 1310\text{nm}$, is equal to $-\pi/2$. The wavelength dependence of a phase shifter $\psi(\lambda)$ can be very well approximated at first order as

$$\frac{d\psi}{\psi} \approx -\frac{d\lambda}{\lambda}. \tag{8}$$

This means that the phase change is proportional to the phase angle and can become important for higher order phase shifters. In particular at $\lambda_0$ we don't need any phase correction, i.e. we must have a null phase shift $\psi_0 = 2n\pi$, where $n$ is an integer to be determined so that $d\psi/d\lambda = -2n\pi/\lambda_0 = -d\phi/d\lambda$. In this way the wavelength change of the additional null phase shifter will compensate for the phase change of the FH MZC. In our numerical example this condition requires $n = -2.4$, that must be approximated by $n = -2$ or $n = -3$, resulting in a non perfectly corrected phase error. Anyway the standard FH MZC, when integrated with this additional phase shifter, effectively form a 50-50 splitter with both stable power and phase output, as required by the criterion for mirror symmetric interferometers.

Figure 8 shows the spectral responses corresponding to the cross ports of three different MZIs. As expected the HFFH configuration and the FHHF configuration with phase correction ($n = -3$ has been chosen) are found to work much better than a simple FHHF structure. In particular the HFFH configuration is flat over a band wider than 70 nm. The geometric representations in Fig. 9 show the trajectories on the sphere for both configurations when the coupling angles are assumed to be 90% of their nominal values.

## Conclusions

Using a general matrix form for 50-50 transformations and with the help of a convenient geometrical representation, we have found a general criterion to design wavelength insensitive mirror-symmetric interferometers. We apply this criterion to wavelength flattened MZCs to design two different implementations of this kind of interferometers.

**Figures**

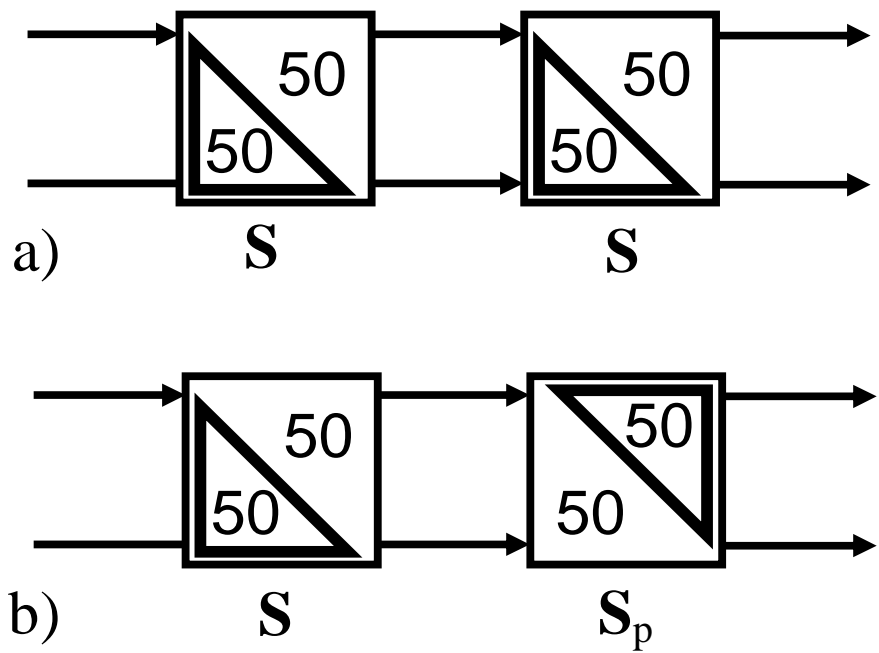

Fig. 1: Schematic of two MZIs: a) composed by two identical 50-50 splitters, each one represented by the matrix **S**, and b) composed by a 50-50 splitter cascaded with its point-symmetric, corresponding to the matrix $\mathbf{S}_p$. We define a "50-50 splitter" any combination of directional couplers and phase shifters giving 50-50 power splitting.

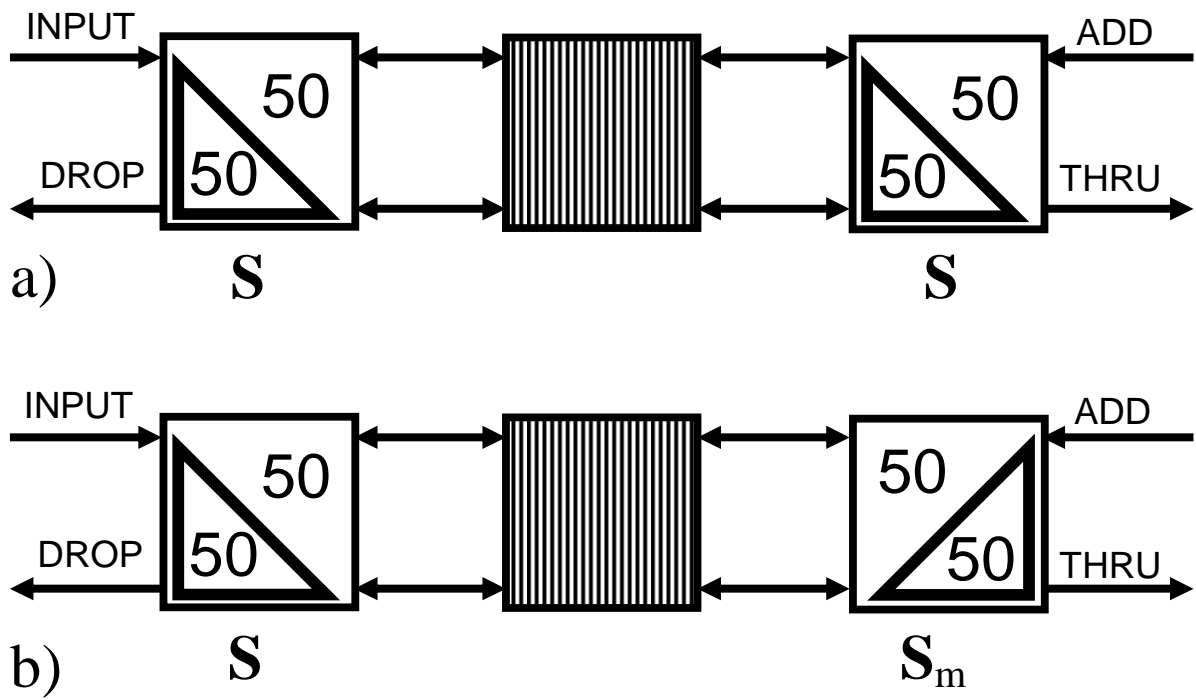

Fig. 2 Schematic of two grating on MZIs: a) composed by two identical 50-50 splitters, each one represented by the matrix **S**, and b) composed by a 50-50 splitter cascaded with its mirror-symmetric, corresponding to the matrix $\mathbf{S}_m$. Also labelled are the four ports of the resulting add/drop filters.

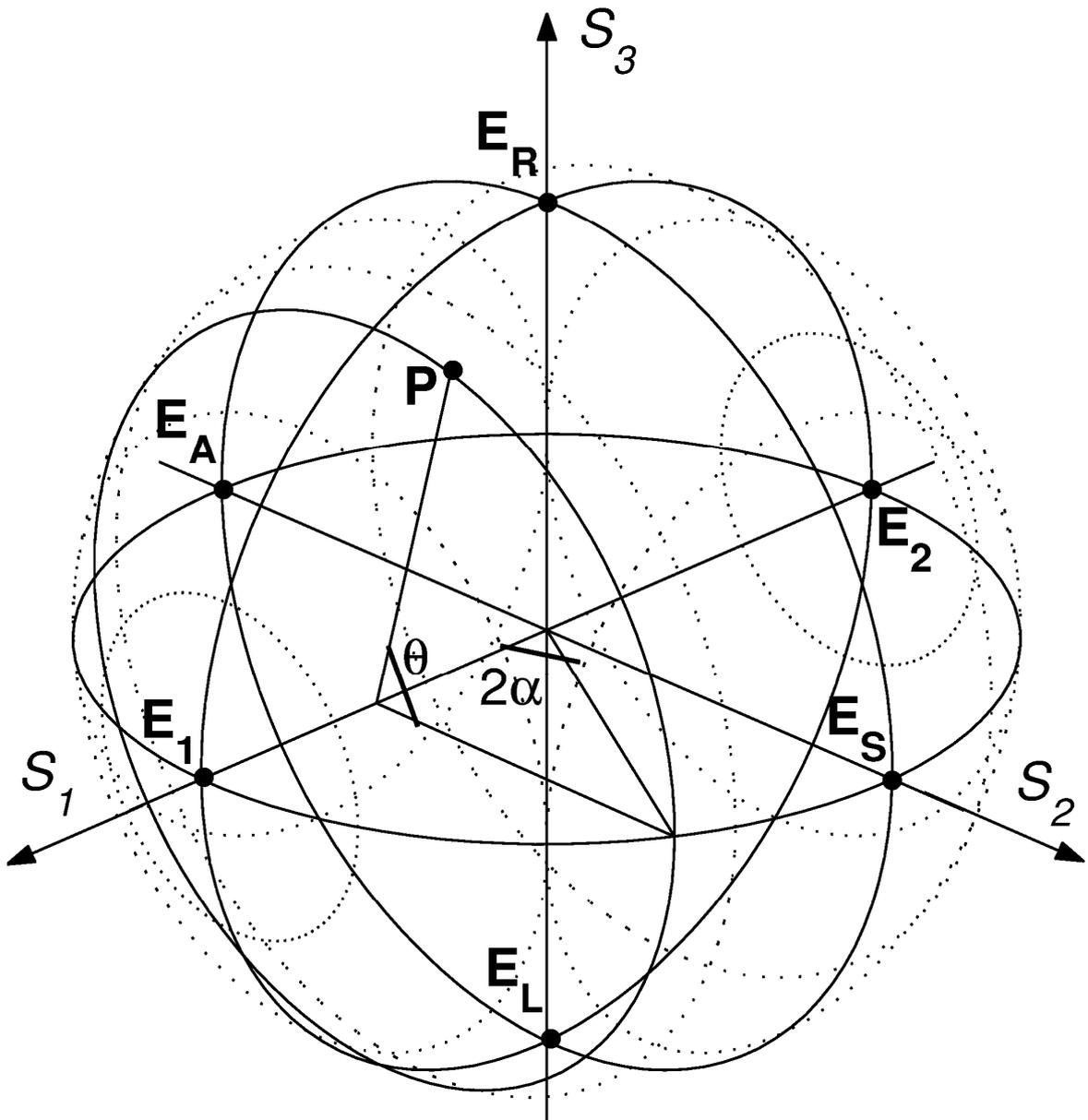

Fig. 3: Generalized Poincaré sphere for the analysis of two coupled waveguides. Physical transformation are represented by composition of rotations about axes on the $S_1S_2$ plane. The points on the rotation axis represent the eigenstates of the system. In particular a synchronous coupler is represented by a rotation about the $S_2$ axis, whereas a phase shifter is represented by a rotation about the $S_1$ axis.

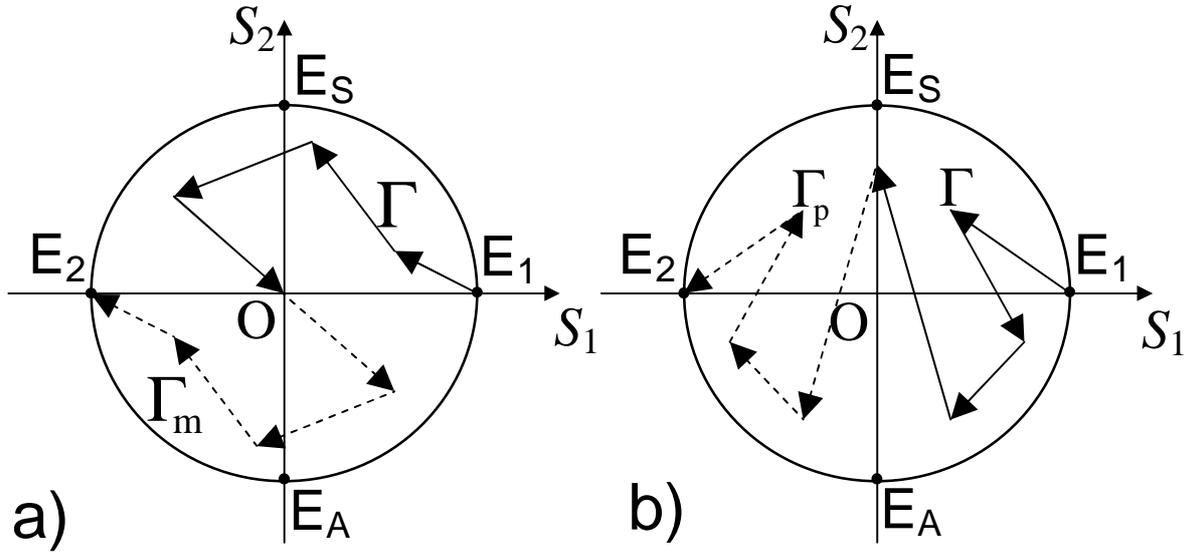

Fig. 4: Projections $\Gamma$ on the $S_1S_2$ plane of the trajectory a) of a 50-50 splitter sending $E_1$ in $E_R$ followed by the projection $\Gamma_m$ of the trajectory corresponding to its mirror-symmetric splitter; and b) of a generic 50-50 splitter cascaded with its point-symmetric splitter, represented by the projection $\Gamma_p$. It is worth to note that the trajectory of a mirror symmetric interferometer is point symmetric with respect to O, whereas the trajectory of a point symmetric interferometer is mirror symmetric with respect to the $S_2$ axis. Physically this means that a point symmetric interferometer can be build with any 50-50 splitter, whereas the 50-50 splitter needed for a mirror symmetric interferometer must pass through $E_R$ or $E_L$ (i.e. the two ports must be in quadrature).

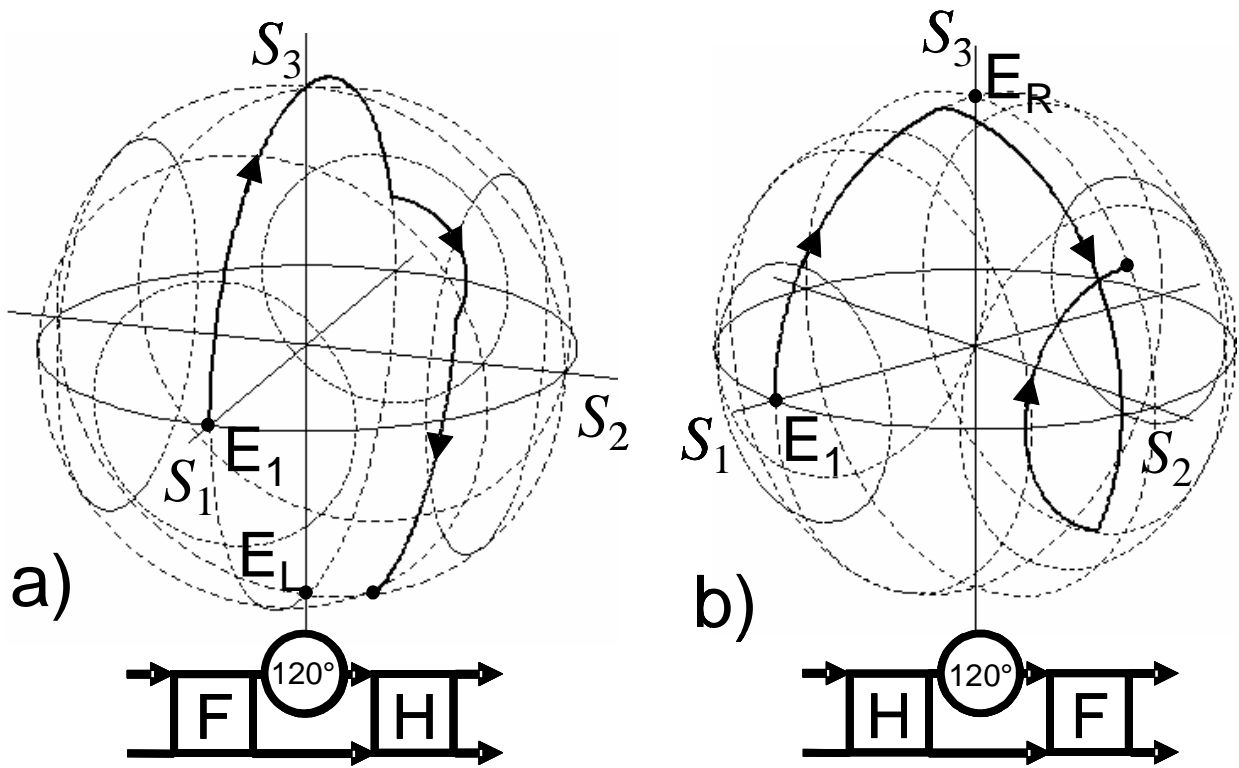

Fig. 5: Geometric representation of the action of a) a FH MZC, and b) a HF MZC. In both cases we have assumed the coupling angles to be 90% of their nominal values to show the insensitivity of the MZC to wavelength changes. The coupling angle error of the full coupler is twice the error of the half coupler, and both are compensated by the fact that $\cos(120°) = -1/2$ (see. text).

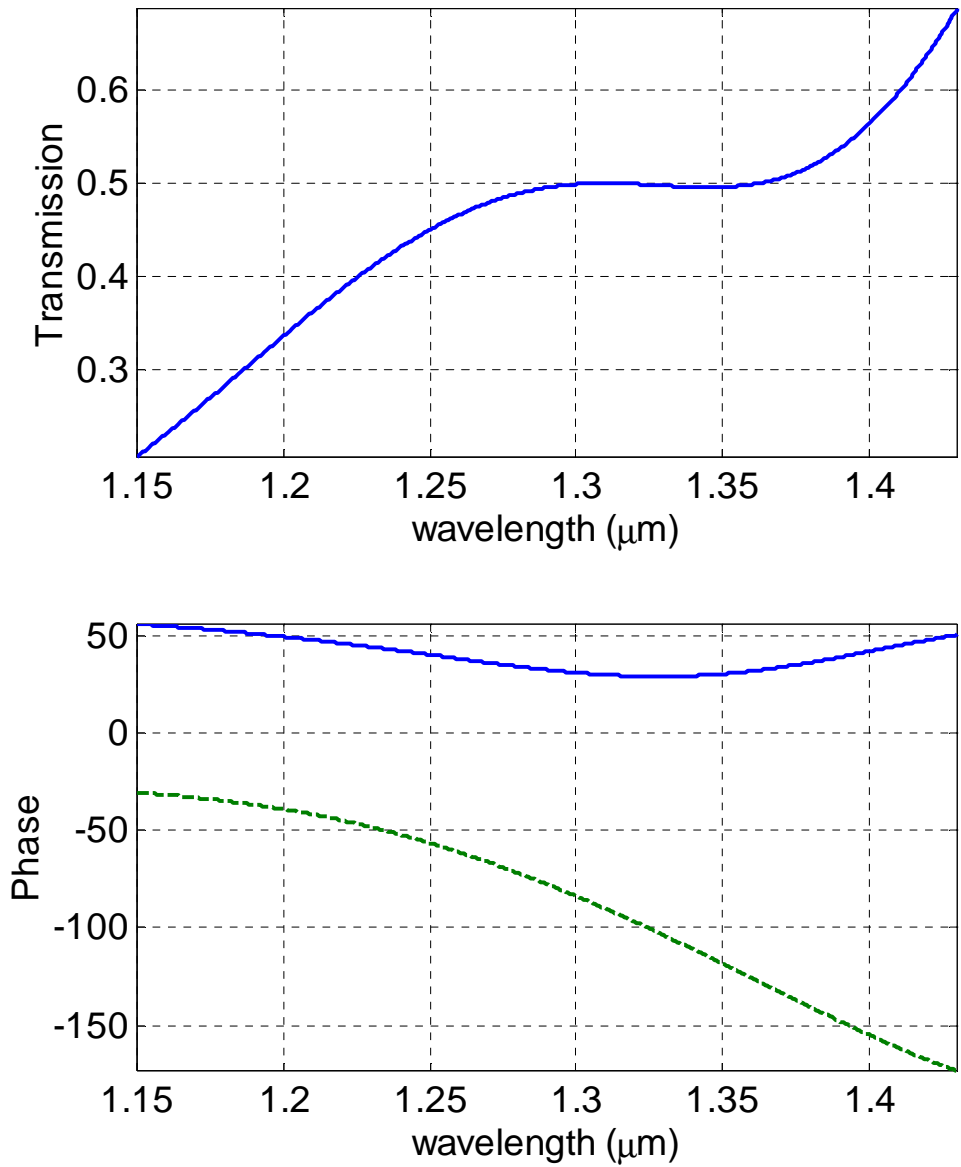

Fig. 6: a) Spectral response for the normalized amplitude in the cross port of a MZC. This is the same both for FH and HF configuration. b) Relative phase between the two arms in the HF MZC (solid line) and in the FH MZC (dashed line). At the nominal wavelength of 1310 nm both configurations feature flat power response, but only the HF MZC ensures also flat phase response.

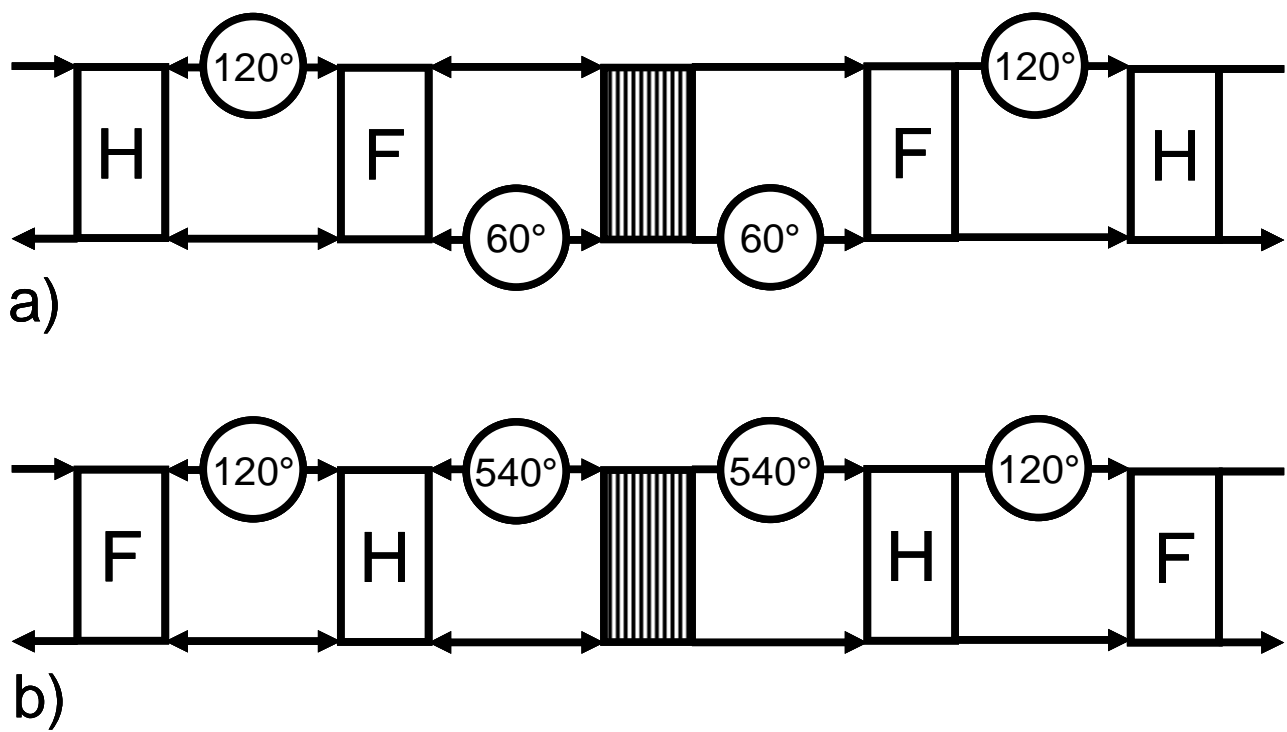

Fig. 7: Wavelength insensitive mirror symmetric MZIs a) in the HFFH configuration and b) in the phase corrected FHHF configuration. In the first case the HF MZC features both stable power and phase response. This means that, in order to design a wavelength insensitive interferometer, we just have to fulfil the criterion for mirror symmetric configurations. Referring to Fig. 5, the simplest way to reach $E_R$ is to cascade an additional 60° phase shifter. In the second case the FH MZC already fulfil the criterion, but its phase response is not stable. A trick to compensate for this wavelength sensitivity is to cascade an higher order nπ phase shifter (in this case n=-3) with opposite phase slope.

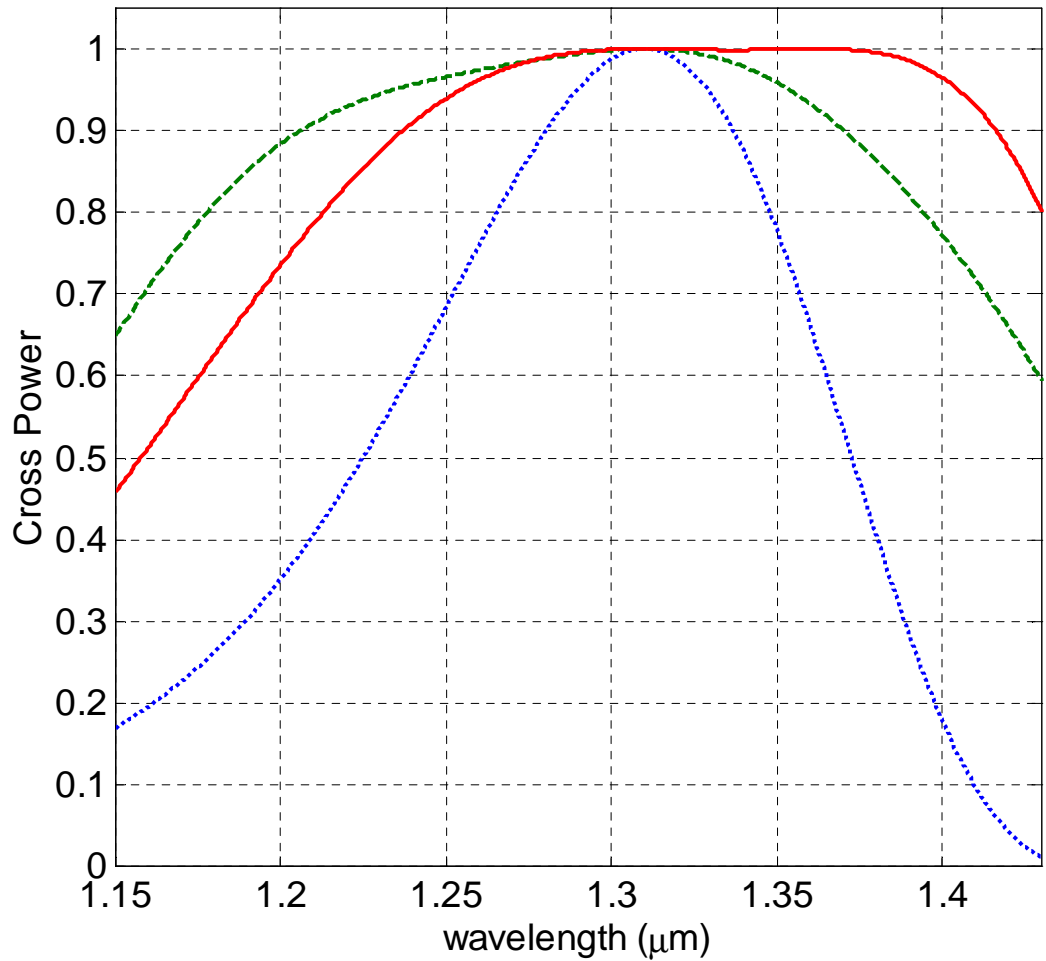

Fig. 8: Spectral response of the normalized power in the cross port of a HFFH MZI (solid line) and of a phase corrected FHHF MZI (dashed line) in comparison with that of a FHHF MZI without phase correction (dotted line). The numerical simulations take into account wavelength dependence of both couplers and phase shifters.

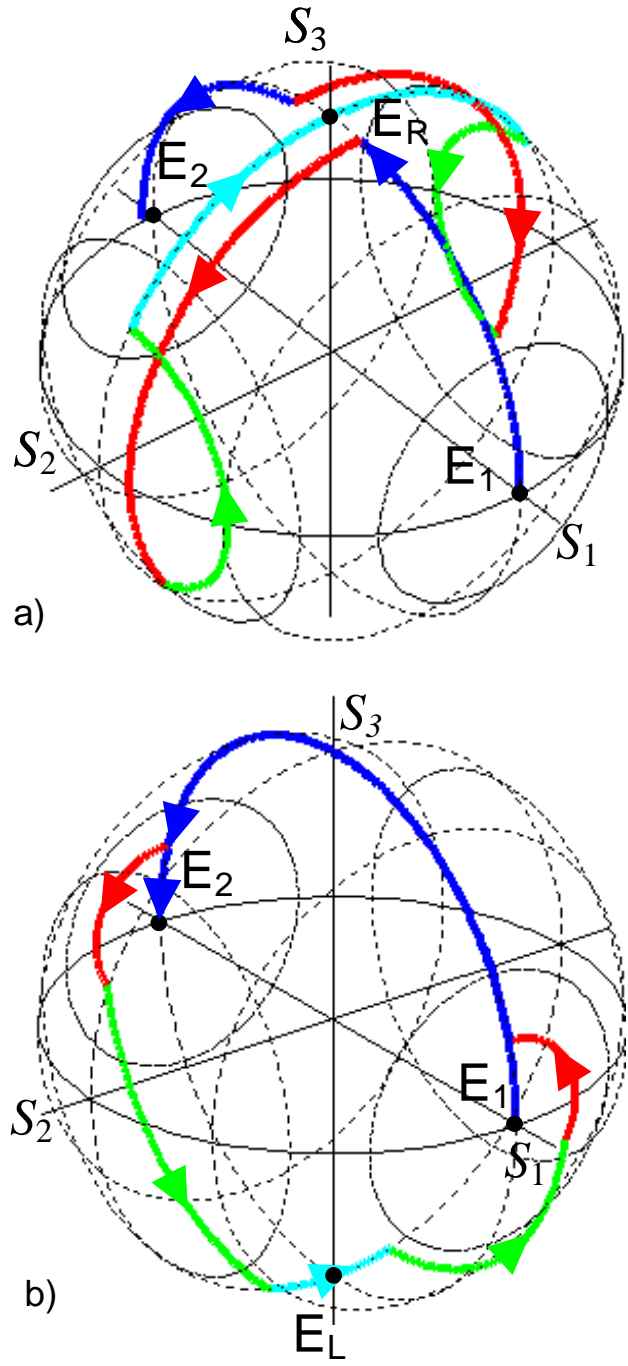

Fig. 9: Geometric representation of the action of the filters in Fig. 7. a) ideally the -120° and the 60° phase shifters should be wavelength independent; b) the 540° phase correction must be wavelength dependent (actually, for the sake of clarity, we have plotted only the phase correction around $E_L$). In both cases we have assumed the coupling angles to be 90% of their nominal values to show the insensitivity to wavelength changes.